\documentclass[letterpaper, 10pt, conference]{ieeeconf}

\usepackage{nopageno}

\IEEEoverridecommandlockouts                             
\usepackage{amsmath}
\usepackage{cite}
\usepackage{bbold}
\usepackage{amssymb}
\usepackage{graphicx}
\usepackage{enumerate}
\usepackage{comment,kotex,color}

\newtheorem{assum}{\bf Assumption}
\newtheorem{rem}{\bf Remark}

\newcommand*{\QEDB}{\hfill\ensuremath{\square}}

\title{\LARGE \bf Open-loop contraction design}
\author{Jin Gyu Lee, Thiago B. Burghi, and Rodolphe Sepulchre
\thanks{This work was partially supported by the National Research Foundation of Korea (NRF) grant funded by the Korea government (Ministry of Science and ICT) (No. NRF-2017R1E1A1A03070342).
The research leading to these results has received partial funding from the European Research Council under the Advanced ERC Grant Agreement Switchlet n.670645.
This work was done while Jin Gyu Lee was with Seoul National University.
}
\thanks{J.~G.~Lee is with Control and Power Research Group, Department of Electrical and Electronic Engineering, Imperial College London, United Kingdom.
T.~B.~Burghi and R.~Sepulchre are with Control Group, Department of Engineering, University of Cambridge, United Kingdom.
{\tt\small jin-gyu.lee@imperial.ac.uk}, {\tt\small tbb29@cam.ac.uk}, {\tt\small r.sepulchre@eng.cam.ac.uk}}
}

\begin{document}

\maketitle
\pagestyle{plain}
\thispagestyle{plain}

\begin{abstract}
Given a non-contracting trajectory of a nonlinear system, we consider the question of designing an input perturbation that makes the perturbed trajectory contracting.
This paper stresses the analogy of this question with the classical question of feedback stabilization. 
In particular, it is shown that the existence of an output variable that ensures contraction of the inverse system facilitates the design of a contracting input perturbation. 
We illustrate the relevance of this question in parameter estimation.
\end{abstract}

\section{Introduction}

Given a non-contracting trajectory of a nonlinear system, we consider the question of designing an input perturbation that makes the perturbed trajectory contracting.
This question has relevance in a number of applications and phenomena:
\begin{itemize}
\item {\bf Open-loop stabilization}: 
In this case, one usually considers an unstable equilibrium point and asks the question of designing an open-loop control that makes the new attractor asymptotically stable.
Examples include Kapitza's pendulum~\cite{kapitza1965dynamical} and planar juggler~\cite{ronsse2007rhythmic}.
\item {\bf Entrainment}:
The question of designing a periodic input trajectory such that the corresponding output trajectory has the same period.
Contraction ensures entrainment~\cite{russo2010global}.
\item {\bf Noise-induced contraction}:
There are experimental and analytical demonstrations that a nonlinear system subject to the appropriate white noise input becomes contractive~\cite{bryant1976spike,mainen1995reliability}.
Induced contraction can be seen as a deterministic version of noise-induced synchronization~\cite{zhou2003noise}.
\end{itemize}

While the question of induced contraction has mostly been studied as a dynamical systems theory question rather than a control question, this paper highlights the close analogy between induced contraction and the classical question of feedback stabilization. 
In particular, the classical paper of Byrnes, Isidori, and Willems observed that feedback stabilization amounts to finding an output function that makes the system stable invertible, namely relative degree one and minimum phase~\cite{bymes1992passivity}. 
We approach the question of induced contraction in the same manner: we observe that the question becomes tractable when an output variable is found that makes the system right invertible with an inverse system that is contracting.
Notable examples include flat systems and conductance-based models of neuronal circuits.

For systems with a contractive inverse, induced contraction can be reformulated as the design of an output perturbation that makes the system contractive.
We will explore a number of approaches (both conventional and new) including linearization-based approaches (utilizing averaging method or differential Lyapunov function) and describing function method, for this latter task.

Unlike feedforward control without stability, control inputs designed in this way are robust even to model uncertainties.

There are many control problems that motivate the question of induced contraction.
In this paper we illustrate an application in nonlinear system identification.
We also refer the reader to~\cite{lee2022induced} for an application in tracking control.

This paper is organized as follows.
In Section~\ref{sec:ic}, we mathematically formalize our question of `induced contraction.'
Then, in Section~\ref{sec:approach}, we explore a number of approaches with corresponding examples.
Application to system identification is made in Section~\ref{sec:application}.
We conclude in Section~\ref{sec:conc}.

\section{Induced contraction}\label{sec:ic}

Even though the problem has room for further generalization, in this note, we consider the system in normal form
\begin{align}\label{eq:non_sys}
\begin{split}
y &= x_1 \,\,\,\,\quad\quad\quad\in  \mathbb{R}, \\
\dot{x}_i &= x_{i+1} \,\,\,\,\,\quad\quad\in \mathbb{R}, \quad i = 1, \dots, r - 1, \\ 
\dot{x}_r &= f(t, x, z, u) \in \mathbb{R}, \quad x  := {\rm col}(x_1, \dots, x_r), \\ 
\dot{z} &= g(t, z, x) \,\quad \in \mathbb{R}^{n-r},
\end{split}
\end{align}
for the simplicity in its notation.
Functions $f$ and $g$ are continuously differentiable with respect to their arguments and the partial derivative $(\partial f/\partial u)(t, x, z, u)$ is uniformly strictly sign definite, hence there exists a function $f_\text{inv}(\cdot, \cdot, \cdot, \cdot)$ such that $f_\text{inv}(t, x, z, v)$ is the unique solution $u$ of the algebraic equation $v = f(t, x, z, u)$.

For the given system~\eqref{eq:non_sys} and the given output reference trajectory $y^*(\cdot)$ with the given control input $u^*(\cdot)$, we consider the situation where around the corresponding output/state reference trajectory ${\rm col}(x^*(\cdot), z^*(\cdot))$, the system~\eqref{eq:non_sys} is non-contracting, i.e., the linear time-varying system obtained by linearizing~\eqref{eq:non_sys} on the reference trajectory,
\begin{align}\label{eq:linearization}
\begin{split}
\begin{bmatrix} \dot{\delta x} \\ \dot{\delta z}\end{bmatrix} &= A^*(t) \begin{bmatrix} \delta x \\ \delta z\end{bmatrix}
\end{split}
\end{align}
is not uniformly asymptotically stable, where $A^*(t)$ is
$$\begin{bmatrix} \begin{bmatrix} 0 & I_{r-1} \end{bmatrix} & 0  \\ \frac{\partial f}{\partial x}(t, x^*(t), z^*(t), u^*(t)) & \frac{\partial f}{\partial z}(t, x^*(t), z^*(t), u^*(t))\\  \frac{\partial g}{\partial x}(t, z^*(t), x^*(t)) & \frac{\partial g}{\partial z}(t, z^*(t), x^*(t))\end{bmatrix}.$$

The question of interest is to design an input perturbation $\Delta u(\cdot)$ such that the new control input $u^{**}(\cdot) := u^*(\cdot) + \Delta u(\cdot)$ makes the perturbed trajectory to be contracting.
That is, the new output/state reference trajectory ${\rm col}(x^{**}(\cdot),z^{**}(\cdot))$ yields a uniformly asymptotically stable linear time-varying system $\dot{\delta \chi} =  A^{**}(t)\delta \chi$, where $A^{**}(t)$ is
$$\begin{bmatrix} \begin{bmatrix} 0 & I_{r-1} \end{bmatrix} & 0  \\  \frac{\partial f}{\partial x}(t, x^{**}(t), z^{**}(t), u^{**}(t)) & \!\! \frac{\partial f}{\partial z}(t, x^{**}(t), z^{**}(t), u^{**}(t))\\  \frac{\partial g}{\partial x}(t, z^{**}(t), x^{**}(t)) & \frac{\partial g}{\partial z}(t, z^{**}(t), x^{**}(t))\end{bmatrix}\!.$$

The main difficulty of this question is in determining $\Delta u(\cdot)$ that induces contraction but is not infinitesimal.
This difficulty disappears under the following assumption.
\begin{assum}\label{assum:inv_con}
The inverse system of~\eqref{eq:non_sys} given as
\begin{align}\label{eq:inv_sys}
\begin{split}
\dot{\bar{z}} &= g(t, \bar{z}, \bar{u})
\end{split}
\end{align}
has the fading memory property~\cite{boyd1985fading}; there exists a class-$\mathcal{K}_\infty$ function $\gamma$ and a decreasing function $ w : \mathbb{R}_{\ge 0} \to (0, 1]$ that converges to zero, so that for any locally essentially bounded, measurable input trajectories, $\hat{u}$, $u$, the solution trajectories $\hat{z}$, $z$ of~\eqref{eq:inv_sys} exists globally and satisfy
$$\left\|\hat{z}(t) - z(t)\right\| \le \gamma\left(\begin{matrix}\sup_{s  \in (-\infty, t)}\end{matrix} \left\|\hat{u}(s) - u(s)\right\| w(t-s)\right)$$
for all $t$. \QEDB
\end{assum}
\begin{rem}
Note that contraction (of the inverse system) as in~\cite{lohmiller1998contraction} for state-space models implies Assumption~\ref{assum:inv_con}. \QEDB
\end{rem}
Under Assumption~\ref{assum:inv_con}, there exists a straightforward method to design a control input $u^{**}(\cdot)$ that makes $A^{**}(\cdot)$ uniformly asymptotically stable. 
The idea is to first design a stationary output/state reference trajectory ${\rm col}(x^{**}(\cdot), z^{**}(\cdot))$ that is contracting, and that satisfies any additional characteristics that we desire, such as small $\sup_{t \in [t_0, \infty)} |y^*(t) - y^{**}(t)|$.
Then, Assumption~\ref{assum:inv_con} guarantees the unique existence of the corresponding input trajectory $u^{**}(\cdot)$ as
$$u^{**}(t) = f_\text{inv}(t, \bar{z}^{**}(t), x^{**}(t), \dot{x}_r^{**}(t)),$$
where $\bar{z}^{**}(\cdot)$ is the stationary trajectory of the inverse system~\eqref{eq:inv_sys} for the input trajectory $\bar{u}(\cdot) = x^{**}(\cdot)$.

In the next section, we will empirically show that many important examples of non-contractive models satisfy Assumption~\ref{assum:inv_con}.
In doing so, we also propose a number of ways to design a new output reference trajectory $y^{**}(\cdot)$.

\begin{rem}
The existence question for such a new output reference trajectory is one of the topics in our future work.
However, we note that if the system has relative degree one, has uniformly strictly sign definite control gain, and satisfies Assumption~\ref{assum:inv_con} as illustrated in the Introduction, then there is at least one output reference trajectory that yields contraction.
In particular, there is a static output feedback $-ky$ that makes the closed-loop system contractive, hence the corresponding stationary output trajectory $y^{**}(\cdot)$ of the closed-loop system induces contraction.
If this output trajectory is identically zero, then we can consider additional rapidly oscillating perturbation as in Section~\ref{subsubsec:aver}. \QEDB
\end{rem}

\section{Approaches to induce contraction}\label{sec:approach}

Based on the discussion given in Section~\ref{sec:ic}, it seems natural to first consider linearization-based approaches.
Depending on the class of models that we are dealing with and depending on the characteristics that we want to impose on our new output reference trajectory, design approaches may vary, where some of them are explored as follows.

\subsection{Open-loop stabilization}\label{subsubsec:aver}

For the initial investigation, we revisit the problem of stabilizing an inverted pendulum given as
\begin{align}\label{eq:Kap}
\begin{split}
\ddot{y} = - \beta \sin(y) - \gamma \dot{y} + \alpha u, \quad \alpha, \beta, \gamma > 0.
\end{split}
\end{align} 
Our goal is to find a perturbation $\Delta y(\cdot)$ that makes the original output reference trajectory $y^*(\cdot) \equiv \pi$ contractive.
For this specific problem, to avoid bias, we require that the perturbation satisfies
$$\int_{-\infty}^\infty \Delta y(t) dt = 0.$$ 
A natural choice is then $\Delta y(t) = M \sin(\omega t)$.
Such choice provides a simple analysis for the stability of the linear time-varying system introduced in Section~\ref{sec:ic}.
In particular, our design reduces to finding an appropriate magnitude $M$ and a frequency $\omega$ such that the linear time-varying matrix
$$A^{**}(t) = \begin{bmatrix} 0& 1 \\ \beta \cos(M \sin(\omega t)) & -\gamma\end{bmatrix}$$
is uniformly asymptotically stable.

\begin{figure}[h]
\begin{center}
\includegraphics[width=\columnwidth]{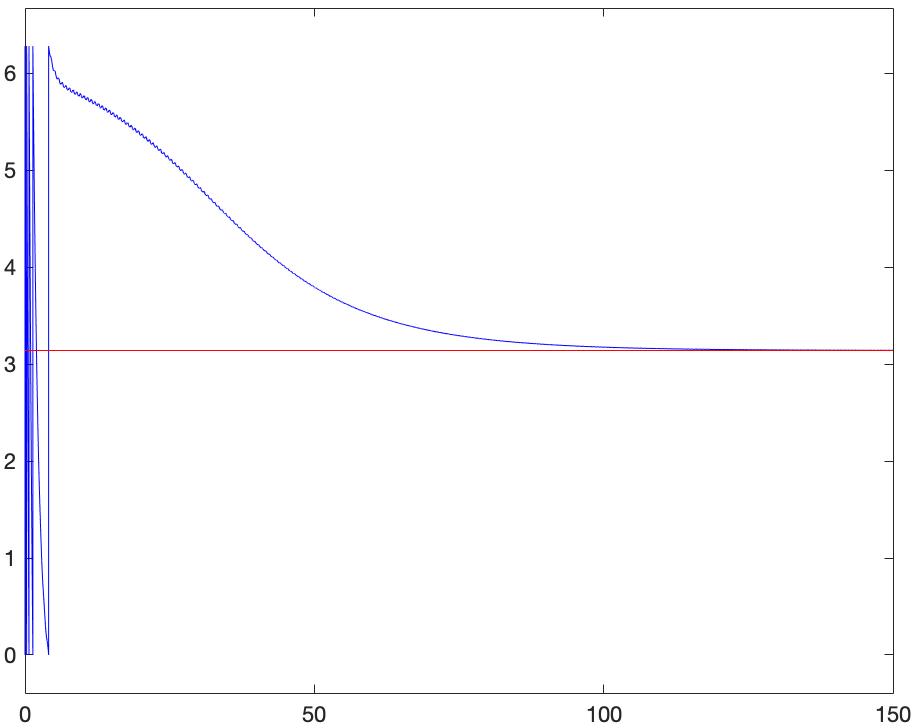}
\caption{Red: $y^*(\cdot)$ and blue: $y(\cdot) - \Delta y(\cdot)$.}
\label{fig:Kap}
\end{center}
\vspace{-.5cm}
\end{figure}

To further simplify our choice, we could assume that this perturbation is rapidly oscillating, i.e., that $\omega \gg 1$.
Then, by averaging, we see that $A^{**}(\cdot)$ is uniformly asymptotically stable for sufficiently large frequency $\omega$ if and only if
$$\overline{A^{**}} := \begin{bmatrix} 0 & 1 \\ \beta\bar{c} & -\gamma\end{bmatrix}$$
is Hurwitz, where 
$$\bar{c} := \frac{\omega}{2\pi} \! \int_0^{2\pi/\omega} \!\!\! \cos(M\sin(\omega t)) dt = \frac{2\omega}{\pi}\! \int_0^M \!\! \frac{\cos(s)}{\omega \sqrt{M^2 - s^2}} ds.$$
Now, $\bar{c}$ is negative, for instance, when $M = 0.8\pi$, and this ensures contraction. 
Figure~\ref{fig:Kap} shows the simulation result with $\alpha = \beta = \gamma =  1$, $M = 0.8\pi$, and $\omega = 1000$.
Note that such a large magnitude in the perturbation $\Delta y$ is necessary if our input gain is constant as in~\eqref{eq:Kap}, in order to visit the region where the Jacobian is Hurwitz (contractive region).

\subsection{Entrainment}

A stable limit cycle is a notable example of solution that is not contractive due to a Floquet multiplier equal to one~\cite{strogatz2018nonlinear}. 
Yet, a stable limit cycle is almost contractive, and ``most'' perturbations of the limit cycle result in a contractive trajectory. 
A popular manifestation of this property is the entrainment phenomenon that a periodic input trajectory with a period close to the period of the limit cycle will result in a contractive periodic trajectory. 
We illustrate the design of such control input on several classical models.

\subsubsection{Impulsive perturbation (Fitzhugh-Nagumo)}\label{subsubsec:imp}

$ $

We first consider entrainment of Fitzhugh-Nagumo oscillators given as ($\epsilon \ll 1$)
\begin{align}\label{eq:FN}
\begin{split}
\epsilon \dot{y} &= \alpha y - \beta y^3 - \gamma z  + u, \quad \alpha, \beta, \gamma > 0,\\
\dot{z} &= -z + y,
\end{split}
\end{align}
which to have a stable limit cycle for sufficiently small relaxation parameter $\epsilon$, the parameters must satisfy $2\alpha < 3\gamma$.
Let us denote a free ($u^*(\cdot) \equiv 0$) stable periodic orbit with a specific initialization as ${\rm col}(y^*(\cdot), z^*(\cdot))$.
Then, the linear time-varying system obtained by linearization of~\eqref{eq:FN} on the original output/state reference trajectory, $\dot{\delta \chi} = A^*(t) \delta\chi$ with
$$A^*(t) := \begin{bmatrix} \frac{1}{\epsilon}\left(\alpha - 3\beta y^*(t)^2\right) & -\frac{\gamma}{\epsilon} \\ 1 & -1\end{bmatrix},$$
satisfies that its state transition matrix denoted by $\Phi^*(t, t_0)$ satisfies that for each $t_0$, the matrix $\Phi^*(t_0 + T^*, t_0)$ has eigenvalue $1$ and $\lambda^* \in (-1, 1)$ (which approaches zero as $\epsilon \to 0$), where $T^* > 0$ is the period of the orbit.

If our objective is to have entrainment with a new output reference trajectory that is almost everywhere identical to the original one, then we can simply choose our output perturbation $\Delta y(\cdot) = y^{**}(\cdot) - y^*(\cdot)$ to be of impulsive nature.

Note that our new linear time-varying system, $\dot{\delta\chi} = A^{**}(t)\delta\chi$ with
\begin{align*}
A^{**}(t) &:= \begin{bmatrix} \frac{1}{\epsilon}\left(\alpha - 3\beta y^{**}(t)^2\right) & -\frac{\gamma}{\epsilon} \\ 1 & -1\end{bmatrix} =: A^*(t) + \Delta A(t), \\
\Delta A(t) &= \begin{bmatrix} -\frac{3\beta}{\epsilon}\left[ y^*(t) + \Delta y(t)\right]\Delta y(t) & 0 \\ 0 & 0 \end{bmatrix} =: \begin{bmatrix} \Delta a(t) & \!0 \\ 0 & \!0 \end{bmatrix}\!,
\end{align*}
has its state transition matrix denoted by $\Phi^{**}(t, t_0)$, which has the relation (by the variation of constants formula)
$$\Phi^{**}(t_0 + T^*, t_0) \equiv \Phi^*(t_0 + T^*, t_0)\Delta \Phi_{t_0}(t_0 + T^*, t_0),$$
for~each~$t_0$,~where~$\Delta \Phi_{t_0}(t, t_0)$~is~the~state~transition~matrix~of
\begin{align*}
\dot{\delta\chi} &= \Phi^*(t, t_0)^{-1}\Delta A(t) \Phi^*(t, t_0)\delta\chi  \\
&= \Delta a(t) \Phi^*(t, t_0)^{-1}\begin{bmatrix} 1 & 0 \\ 0 & 0 \end{bmatrix} \Phi^*(t, t_0)\delta\chi.
\end{align*}
So, to ease the analysis, let us choose our output perturbation as a train of impulses:
\begin{align}\label{eq:imp_train}
\Delta y(t) = \sum_{n=0}^\infty
\epsilon_n\sqrt{\delta(t - t_0 - nT^*)}.
\end{align}
Then, we see that the new state transition matrix satisfies
\begin{align*}
&\Phi^{**}(t_0 + (n+1)T^*, t_0 + nT^*) \\
&\quad\quad\quad\quad =\Phi^*(t_0 + (n+1)T^*, t_0 + nT^*)\begin{bmatrix} 1 - \frac{3\beta}{\epsilon}\epsilon_n^2 & 0 \\ 0 & 1\end{bmatrix} \\
&\quad\quad\quad\quad =\Phi^*(t_0 + T^*, t_0)\begin{bmatrix} 1 - \frac{3\beta}{\epsilon}\epsilon_n^2 & 0 \\ 0 & 1\end{bmatrix}.
\end{align*}
Hence, we further choose $\epsilon_n  > 0$ to satisfy $3\beta \epsilon_n^2 < \epsilon$.

Now, if we denote the right (left) eigenvector of the matrix $\Phi^*(t_0 + T^*, t_0)$ associated with the eigenvalue $1$ and $\lambda^*$ as $\mathsf{v}$ ($\bar{\mathsf{v}}$) and $\mathsf{w}$ ($\bar{\mathsf{w}}$) respectively, then
$$\Phi^*(t_0 + T^*, t_0) = \begin{bmatrix} \mathsf{v} & \mathsf{w}\end{bmatrix} \begin{bmatrix} 1 & 0 
\\ 0 & \lambda^* \end{bmatrix} \begin{bmatrix} \bar{\mathsf{v}}^T \\ \bar{\mathsf{w}}^T\end{bmatrix},$$
and the matrix $\Phi^{**}(t_0 + (n+1)T^*, t_0 + nT^*)$ becomes stable (i.e., all the eigenvalues are contained inside the unit circle), and thus contraction is induced, when and only when
$$\begin{bmatrix} 1 & 0 \end{bmatrix} \mathsf{v} \neq 0.$$
Therefore, our design problem reduces to finding an instant $t_0 \in [0, T^*)$ such that $\begin{bmatrix} 1 & 0 \end{bmatrix}\mathsf{v}(t_0) \neq 0$.
Then, a train of impulses~\eqref{eq:imp_train} can be utilized to induce contraction.

\begin{figure}[h]
\begin{center}
\includegraphics[width=\columnwidth]{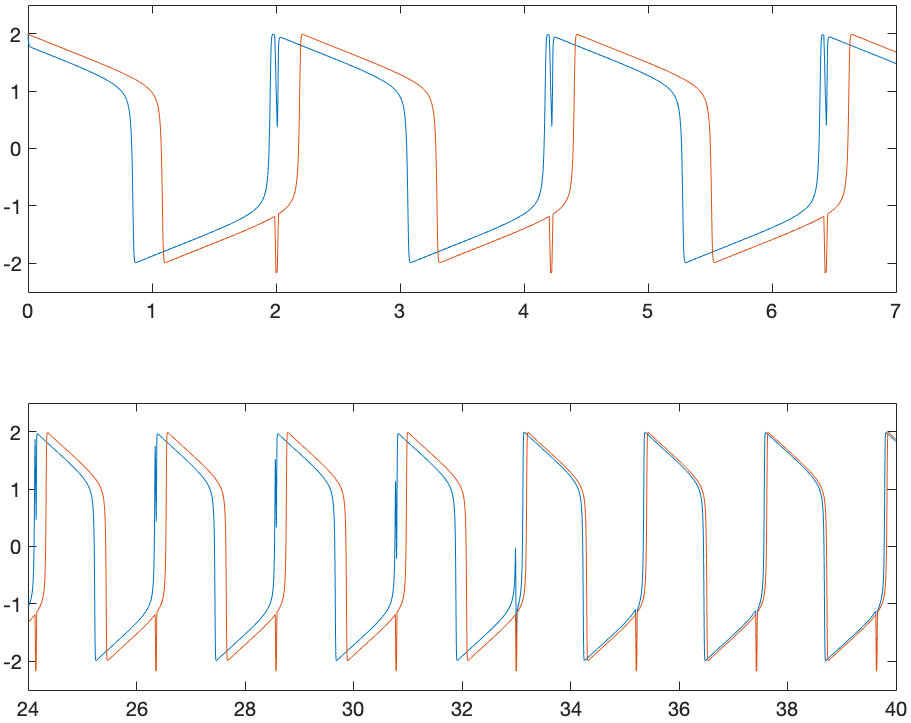}
\caption{Red: $y^{**}(\cdot)$ and blue: $y(\cdot)$.}
\label{fig:FN}
\end{center}
\vspace{-.5cm}
\end{figure}

Since, we can find an explicit representation of the right eigenvector $\mathsf{v}(t_0)$ as
$$\mathsf{v}(t_0) = \begin{bmatrix} \frac{1}{\epsilon}\left[\alpha y^*(t_0) - \beta y^*(t_0)^3 - \gamma z^*(t_0)\right]\\ -z^*(t_0) + y^*(t_0)\end{bmatrix},$$
because $\mathsf{v}(t_0)$ is simply the tangential direction of the limit cycle at point ${\rm col}(y^*(t_0), z^*(t_0))$, almost any instant $t_0 \in [0, T^*)$ can be chosen as a suitable design parameter.
Note that, if this is synchronized to the instant where the jump happens, the effect becomes maximized.
Figure~\ref{fig:FN} shows the simulation result with $\alpha = \beta = \gamma = 1$, where the square of the Dirac delta function is realized by its approximation:
$$\sqrt{\delta(x)} = \lim_{a \to 0} \sqrt{\frac{1}{|a|\sqrt{\pi}}e^{-(x/a)^2}}.$$

\subsubsection{Differential Lyapunov function (Hodgkin-Huxley-type)}\label{subsubsec:Lya}

$ $

\vspace{-4mm}
Now, we extend our interest to general conductance-based models of form
\begin{align}\label{eq:HH}
\begin{split}
\epsilon\dot{y} &= -g\!\left(y \! - \! E\right)  - \bar{g}_f\!\left[1 \!+\! \tanh\!\left(\kappa_f\!\left(y \!-\! V_f\right)\right)\right]\!\left(y \!-\! E_f\right) \\
&\quad - \bar{g}_s \!\left[1\! + \!\tanh\!\left(\kappa_s\!\left(z \!-\! V_s\right)\right)\right]\!\left(y \!-\! E_s\right) + u, \\
\dot{z} &= -z + y,
\end{split}
\end{align}
where $g, \bar{g}_f, \bar{g}_s, E, E_f, E_s, V_f$, and $V_s$ are design parameters such that $g, \bar{g}_f, \bar{g}_s > 0$ and $E_s < E, V_f, V_s < E_f$.

For the problem of designing a periodic input trajectory that yields entrainment, let us consider a new output/state reference trajectory ${\rm col}(y^{**}(\cdot), z^{**}(\cdot))$ that is $T^{**}$-periodic.
Then, the linear time-varying system obtained by linearization of~\eqref{eq:HH} on this trajectory can be found as $\dot{\delta\chi} = A^{**}(t)\delta\chi$, where
\begin{align*}
A^{**}(t) &:= \begin{bmatrix} -g_\text{tot}^{**}(t)/\epsilon & -g_s^{**}(t)/\epsilon \\ 1 & -1 \end{bmatrix}, \\
g_\text{tot}^{**}(t) &:= g + \bar{g}_f\left[1 + \tanh\left(\kappa_f\left(y^{**}(t) - V_f\right)\right)\right]\\
&\quad + \bar{g}_s\left[1 + \tanh\left(\kappa_s\left(z^{**}(t) - V_s\right)\right)\right] \\
& + \bar{g}_f\kappa_f\!\left[1 - \tanh^2\!\left(\kappa_f\!\left(y^{**}(t) - V_f\right)\right)\right]\!\left(y^{**}(t) - E_f\right)\!,\\
g_s^{**}(t) &:= \bar{g}_s\kappa_s \! \left[1 - \tanh^2\!\left(\kappa_s\!\left(z^{**}(t) - V_s\right)\right)\right]\!\left(y^{**}(t) - E_s\right)\!.
\end{align*}
Now, a sufficient condition for this $T^{**}$-periodic linear time-varying system to be uniformly asymptotically stable can be found by observing the differential Lyapunov function 
$$\delta V := \frac{1}{2}\frac{\epsilon}{g_s^{**}(t)}\delta y^2 + \frac{1}{2}\delta z^2.$$
In particular, its time derivative can be found as
$$\dot{\delta V} = -\frac{g_\text{tot}^{**}(t)}{g_s^{**}(t)}\delta y^2 - \frac{1}{2}\frac{\epsilon\dot{g}_s^{**}(t)}{g_s^{**}(t)^2}\delta y^2 - \delta z^2.$$
Note that
$$\frac{\dot{g}_s^{**}(t)}{g_s^{**}(t)} = \frac{\dot{y}^{**}(t)}{y^{**}(t) - E_s} - 2\kappa_s\dot{z}^{**}(t)\tanh\left(\kappa_s\left(z^{**}(t) - V_s\right)\right),$$
hence if we assume that $z^{**}(t),y^{**}(t) \in [E_s + \theta, E_f - \theta']$ and $|\epsilon\dot{y}^{**}(t)| \le M_y$ for all $t  \in [0,  T^{**})$ (with some $\theta, \theta', M_y > 0$), then we get
\begin{align*}
\left|\frac{1}{2}\frac{\epsilon\dot{g}_s^{**}(t)}{g_s^{**}(t)}\right| &\le \frac{M_y}{2\theta} + \epsilon\kappa_s(E_f - E_s) =: M_s^{**}.
\end{align*}
Also, we get
\begin{align*}
\left|g_\text{tot}^{**}(t)\right| \le  g + 2\bar{g}_f + 2\bar{g}_s + \bar{g}_f\kappa_f(E_f - E_s) =: G_\text{tot}^{**}.
\end{align*}

This implies that $\dot{\delta V} \le 2(\bar{a}/\epsilon)\delta V$ if 
\begin{align}\label{eq:unstable_cond}
-g_\text{tot}^{**}(t) - \frac{1}{2}\frac{\epsilon\dot{g}_s^{**}(t)}{g_s^{**}(t)} > 0,
\end{align}
where $\bar{a} := M_s^{**} + G_\text{tot}^{**}$, $\dot{\delta V} \le -2\delta V$ if
\begin{align}\label{eq:stable_cond}
-g_\text{tot}^{**}(t) - \frac{1}{2}\frac{\epsilon\dot{g}_s^{**}(t)}{g_s^{**}(t)} \le -\epsilon,
\end{align}
and $\dot{\delta V} \le 0$ otherwise.
So, during one period of time, if the measure of the time interval corresponding to the condition~\eqref{eq:unstable_cond} is smaller than $\tau^{**} > 0$ and if the measure of the time interval corresponding to the condition~\eqref{eq:stable_cond} is larger than $\hat{T}^{**}$, then we can conclude that 
$$\delta V(t_0 + T^{**}) \le e^{-2\hat{T}^{**}}e^{2(\bar{a}/\epsilon)\tau^{**}}\delta V(t_0).$$
Hence, a sufficient condition for contraction is $\epsilon\hat{T}^{**} > \bar{a}\tau^{**}$. 

\begin{figure}[h]
\begin{center}
\includegraphics[width=\columnwidth]{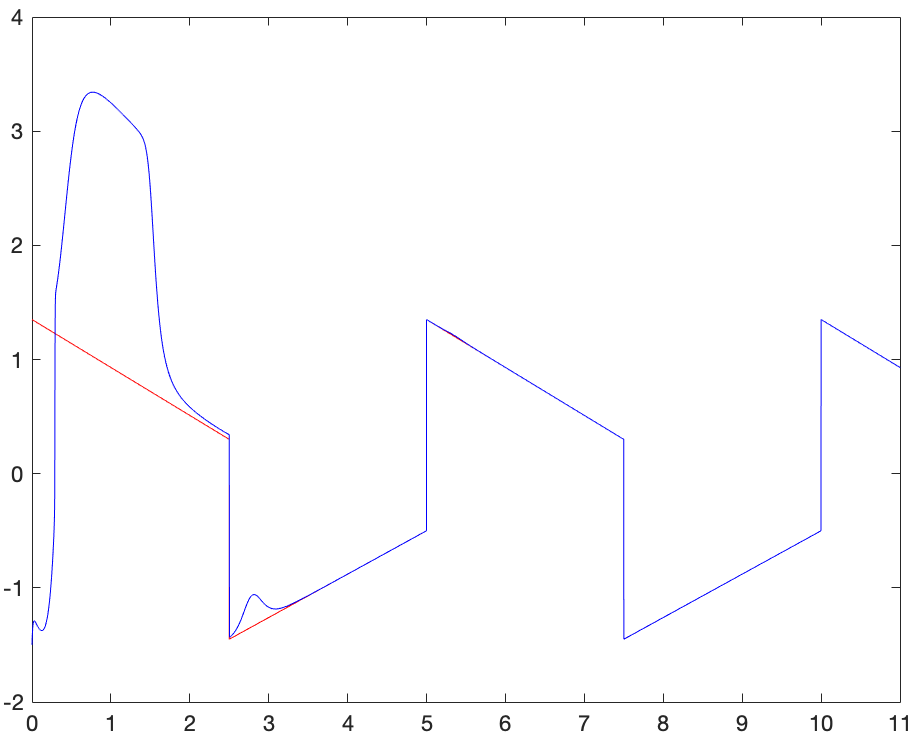}
\caption{Red: $y^{**}(\cdot)$ and blue: $y(\cdot)$.}
\label{fig:HH}
\end{center}
\vspace{-.5cm}
\end{figure}

In particular, if we consider the system
\begin{align*}
0.01\dot{y} &= -y - 2\left[1 + \tanh(5y)\right](y - 2) \\
&\quad - 2\left[1 + \tanh(5z)\right](y + 2) + u,  \\
\dot{z} &= - z + y,
\end{align*}
with $M_y = 0.33$, $\theta  = 0.55$, and $\theta' = 0.65$, then~\eqref{eq:stable_cond} is satisfied if $y^{**}(t) \in [-1.45, 1.35]\setminus [-0.5, 0.3] =: \mathcal{I}$ because
\begin{align*}
g_\text{tot}^{**}(t) &\ge 1 + 2 + 2 - 2*5*\begin{matrix}\sup_{y \in \mathcal{I}}\end{matrix}[1 - \tanh^2(5y)](2 - y)\\
& \ge 5 - 3.1 \ge 0.51 =  0.01 + \frac{0.33}{1.1} + 0.01*5*4 \\
&= \epsilon + M_s^{**} \ge \epsilon -  \frac{1}{2}\frac{\epsilon\dot{g}_s^{**}(t)}{g_s^{**}(t)}.
\end{align*}
Therefore, if the new output reference trajectory stays in $\mathcal{I}$ for a long enough time $\hat{T}^{**}$, then it becomes contractive.
Figure~\ref{fig:HH} shows the simulation result with the square wave-like output reference trajectory $y^{**}(\cdot)$ given as
$$\begin{cases} 1.35 - \frac{2.1}{\hat{T}^{**}}t, &\mbox{ if } t \in \left[0, \frac{\hat{T}^{**}}{2}\right], \\ 0.3 - \frac{3.5}{\tau}\left[t - \frac{\hat{T}^{**}}{2}\right], &\mbox{ if } t \in\left[\frac{\hat{T}^{**}}{2}, \frac{T^{**}}{2}\right], \\ -1.45 + \frac{1.9}{\hat{T}^{**}}\left[t - \frac{T^{**}}{2}\right], &\mbox{ if } t \in \left[\frac{T^{**}}{2}, \frac{T^{**} + \hat{T}^{**}}{2}\right], \\ -0.5 + \frac{3.7}{\tau}\left[t - \frac{T^{**} + \hat{T}^{**}}{2}\right], &\mbox{ if } t \in \left[\frac{T^{**} + \hat{T}^{**}}{2},T^{**}\right], \end{cases}$$
where $\hat{T}^{**} = 5$, $\tau = 0.001$, and $T^{**} = \hat{T}^{**} + \tau$.
Note that, we have $\bar{a} = 49.5$, hence
$$\epsilon\hat{T}^{**} = 0.05 > 0.0495 = \bar{a}\tau > \bar{a}\tau^{**}.$$

Motivated by this, if our goal is to have entrainment with an overall small perturbation in the output; if our objective is to have a small supremum norm on $|y^*(\cdot) - y^{**}(\cdot)|$, where ${\rm col}(y^*(\cdot), z^*(\cdot))$ is a free ($u^*(\cdot) \equiv 0$) stable periodic orbit of~\eqref{eq:HH}, then we can simply choose our new output/state reference trajectory as 
$${\rm col}(y^{**}(\cdot), z^{**}(\cdot)) := (1 + \delta) {\rm col}(y^*(\cdot), z^*(\cdot)),$$
with an appropriately small but strictly positive design parameter $\delta$.
In particular, the supremum norm on $|y^*(\cdot) - y^{**}(\cdot)| = \delta|y^*(\cdot)|$ can be made arbitrarily small, and also the input perturbation thus utilized, i.e., $\lim_{\delta \to 0} \Delta u(t) = 0$.
By this scaling of the periodic orbit, the new output/state reference trajectory spends more time on the contractive region than the original output/state reference trajectory (which is marginally stable), hence results in contraction.

\begin{rem}\label{rem:cont}
Again motivated by this, if we consider a specific conductance-based model which has a single slow branch and a spike that results in a specific form of fast-slow oscillation (see Section~\ref{sec:application} for the detailed model), then an input perturbation that induces contraction can be directly found by empirical observation.
In particular, a periodic train of negative impulses as a control input will be sufficient, if its period is smaller than the period of the original fast-slow oscillation.
This is because, by the condition on the period, at least one impulse acts during a single period of oscillation, and it acts in a way to provide time lag, which means that it spends more time on the contractive region. 
An idea to utilize this robust way of inducing contraction in system identification is illustrated in Section~\ref{sec:application}.
\QEDB
\end{rem}

\begin{rem}
We emphasize that these approaches are not limited to handling unstable equilibrium points or limit cycles.
For example, the Lorenz system, which has a strange attractor, given as
\begin{align*}
\dot{{x}}_1 &= \sigma ({x}_2 - {x}_1), \\
\dot{{x}}_2 &= {x}_1(\rho - z) - {x}_2 + u, \\
\dot{z} &= {x}_1{x}_2 - \beta z
\end{align*}
can similarly be handled to find new output trajectory ${x}_1^{**}(\cdot) = y^{**}(\cdot)$ that makes the system contractive.
The linearization results in a time-varying matrix
$$A^{**}(t) = \begin{bmatrix} -\sigma & \sigma & 0 \\ \rho - z^{**}(t) & -1 & -x_1^{**}(t) \\ x_2^{**}(t) & x_1^{**}(t) & -\beta \end{bmatrix}$$
and we only have to draw our new reference trajectory to span much time on the contractive region, e.g., $|\sigma + \beta - z| < 2\sqrt{\sigma/2}$ and $|x_2| < 2\sqrt{\sigma\beta/2}$.
This region is contractive for the differential Lyapunov function $\delta x_1^2 + \delta x_2^2 + \delta z^2$. \QEDB
\end{rem}

\subsubsection{Describing function approach (Lure systems)}\label{subsec:Desc}

$ $

We finally consider entrainment of Lure systems given as
\begin{align}\label{eq:Lure}
\begin{split}
\dot{x} &= Ax + B\left[u - h(y)\right], \\
y &= Cx.
\end{split}
\end{align}
whereby restricting the form of new output reference trajectory as sinusoidal, 
$$y^{**}(t) = M \sin(\omega t),$$
making ease of analysis by utilizing the describing function approach to handle nonlinearity $h(\cdot)$ by bringing the analysis to linear theory of transfer function.

For this purpose, we assume that the linear system $P(s) = C(sI - A)^{-1}B$ has the low-pass filter property.
Then, any high-frequency term of $h(y^{**}(\cdot))$ will vanish, and thus, we can approximate $h(\cdot)$ by the corresponding describing function $H(M, \omega)(s)$ defined as
$$H(M, \omega)(s) = p(M, \omega) + q(M, \omega)s,$$
where
\begin{align*}
p(M, \omega) &= \frac{1}{\pi M} \int_0^{2\pi/\omega} h(M\sin(\omega t))\sin(\omega t) dt, \\
q(M, \omega) &= \frac{1}{\pi M \omega} \int_0^{2\pi/\omega} h(M \sin(\omega t)) \cos(\omega t) dt.
\end{align*}

Then, we can simply look for $(M, \omega)$ such that the closed-loop linear system
$$G(s) = \frac{P(s)}{1+ P(s)H(M, \omega)(s)}$$
is stable.
This is sufficient if $(M, \omega)$ makes the open-loop linear system $H(M, \omega)(s)P(s)$ passive.

Given the choice of parameters $(M, \omega)$, we can find the corresponding input trajectory as the input for $G(s)$ that has its steady-state response as $y^{**}(\cdot)$, i.e., $u^{**}(t) = D \sin(\omega t + \theta)$ such that
\begin{align*}
M\sin(\omega t) = g(\omega) D \sin(\omega t + \theta) + \omega f(\omega) D\cos(\omega t + \theta)
\end{align*}
where $G(j\omega) = g(\omega)  + j\omega  f(\omega)$.
To increase the precision, we can use multi sine input perturbation to approximate the high-order terms of the static nonlinearity.

\begin{figure}[h]
\begin{center}
\includegraphics[width=\columnwidth]{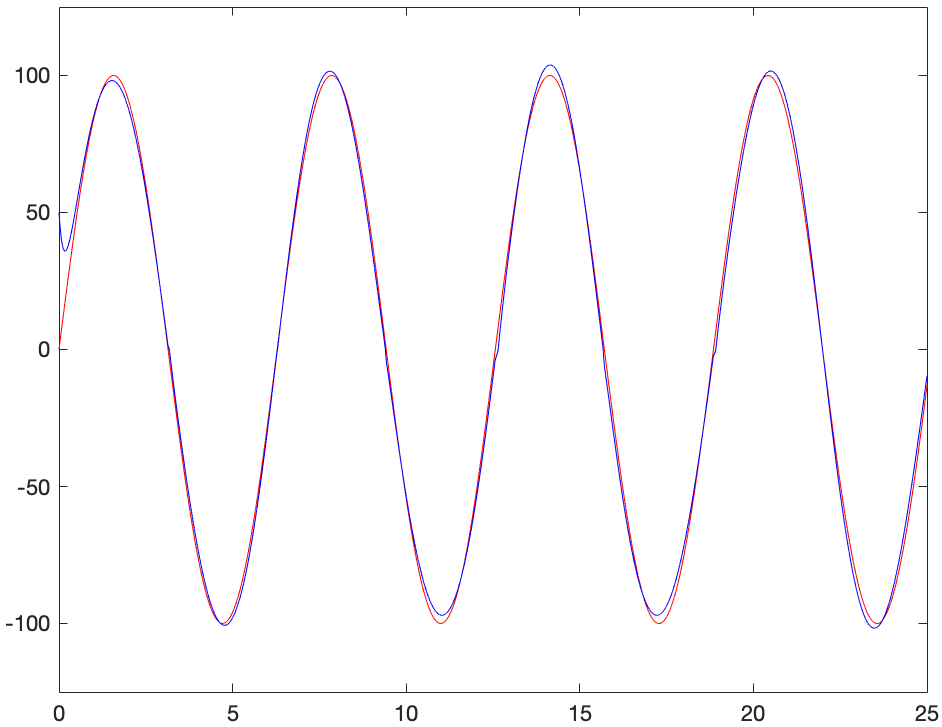}
\caption{Red: $y^{**}(\cdot)$ and blue: $y(\cdot)$.}
\label{fig:Lure}
\end{center}
\vspace{-.5cm}
\end{figure}

For instance, if we consider a Chua circuit, where
$$P(s) = \frac{2s^2 + 0.7s  + 7}{0.2 s^3 + 1.47 s^2 + 0.7 s + 4.9}$$
and
$$h(y) = \begin{cases} -0.1(y + 1) + 4, &\mbox{ if } y \le -1, \\ -4y, &\mbox{ if } -1 < y < 1, \\ -0.1(y-1) - 4, &\mbox{ if } y \ge 1,\end{cases}$$
then we can find the describing function as
\begin{align*}
H(M, \omega)(s) &= p(M,\omega) \\
&= -\frac{7.8}{\pi\omega}\left[\sin^{-1}\left(\frac{1}{M}\right) + \sqrt{\frac{1}{M^2} - \frac{1}{M^4}}\right]
\end{align*}
if $M > 1$ and $H(M, \omega)(s) = p(M, \omega) = -4/\omega$ if $M \le 1$.
Now, since the closed-loop transfer function
$$\frac{P(s)}{1 + \rho P(s)}$$
is stable for $\rho \ge -0.05$, any $(M, \omega)$ such that $H(M, \omega) \in (-0.05,0)$ will suffice.
Example choices are $(M, \omega) = (200, 1)$ and $(M, \omega)= (10, 10)$.
Figure~\ref{fig:Lure} shows the simulation result with the first choice and $u^{**}(t)  = D\sin(\omega t + \theta) + h(M\sin(\omega t)) - p(M, \omega)M\sin(\omega t)$, to compensate for the high-order terms of $h(\cdot)$.

\section{Application: Adaptive observer}\label{sec:application}

Now, we find an application of induced contraction in the problem of parameter estimation.
For this purpose, consider a parameterized family of systems given as
\begin{align}\label{eq:non_sys_app}
\begin{split}
\dot{y} &= f(t, y, z, u) + h(y)^T\theta \in \mathbb{R},  \\ 
\dot{z} &= g(t, z, y) \,\,\quad\quad\quad\quad\quad \in \mathbb{R}^{n-1}.
\end{split}
\end{align}
We assume that there is a uniform input trajectory $u^{**}(\cdot)$ that induces contraction to the system~\eqref{eq:non_sys_app} for all parameters $\theta$ in a given set $\Theta \subset \mathbb{R}^m$.
This control input could be found for instance by any of the methods previously discussed. 
Here, we assume that as in Section~\ref{subsubsec:Lya} that there exists a differential Lyapunov function of form
$$\delta V_\theta = \frac{1}{2}{\delta y}^2 + \frac{1}{2}\delta z^TP_\theta(t)\delta z$$
that decreases along with the new parameterized reference output/state trajectory ${\rm col}(y_\theta^{**}(\cdot), z_\theta^{**}(\cdot))$ during one period of time $T_\theta^{**}$, i.e., $\delta V_\theta (t_0 + T_\theta^{**}) < \delta V_\theta(t_0)$.

Then, by considering $\theta$ also as a state variable, we can construct an adaptive observer that estimates parameters in real-time, by inducing contraction, as
\begin{subequations}\label{eq:non_sys_obs}
\begin{align}
\dot{\hat{y}} &= f(t, \hat{y}, \hat{z}, u^{**}(t)) + h(\hat{y})^T\hat{\theta} \in \mathbb{R},\label{eq:non_sys_obs_a} \\ 
\dot{\hat{z}} &= g(t, \hat{z}, \hat{y}) \,\,\quad\quad\quad\quad\quad\quad\quad \in \mathbb{R}^{n-1}, \label{eq:non_sys_obs_b}\\
\dot{\hat{\theta}} &= -H(\hat{y}) + H(y) \,\,\,\quad\quad\quad\quad  \in\mathbb{R}^m,\label{eq:non_sys_obs_c}
\end{align}
\end{subequations}
where $H(y) = \int h(y) dy$.
We remark that throughout the adaptive observer literature (e.g.,~\cite{besanccon2000remarks,farza2009adaptive,burghi2021online}), the measured output $y$ is injected in the vector field of the adaptive observer, instead of $\hat{y}$ in~\eqref{eq:non_sys_obs_a}-\eqref{eq:non_sys_obs_b}. 
The advantage of this so-called output injection is to facilitate the convergence of the adaptive observer. 
Here, we do not resort to output injection and keep the terms with $\hat{y}$ as can be seen on the right-hand side of~\eqref{eq:non_sys_obs_a}-\eqref{eq:non_sys_obs_b}. 
This particular design choice mitigates potential measurement noise in $y(\cdot)$, while the convergence of the observer will be enabled by induced contraction.

\begin{figure}[h]
\begin{center}
\includegraphics[width=\columnwidth]{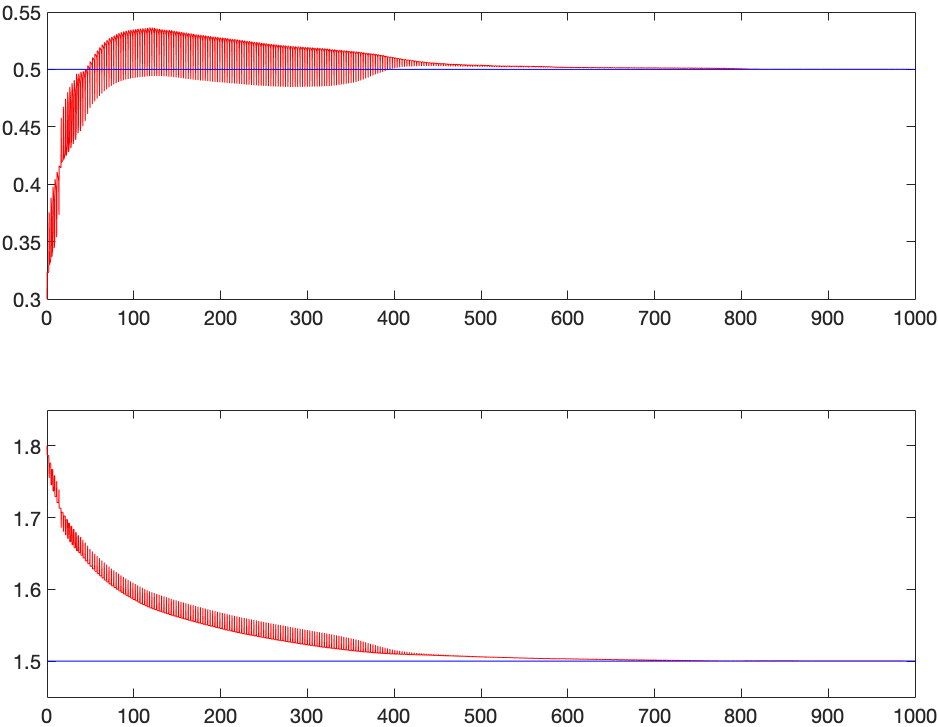}
\caption{Red: $\hat{\theta}(\cdot)$ and blue: $\theta^{**}$, where $\hat{\theta}(0) = {\rm col}(0.3, 1.8)$.}
\label{fig:obs1}
\end{center}
\vspace{-.5cm}
\end{figure}

\begin{figure}[h]
\begin{center}
\includegraphics[width=\columnwidth]{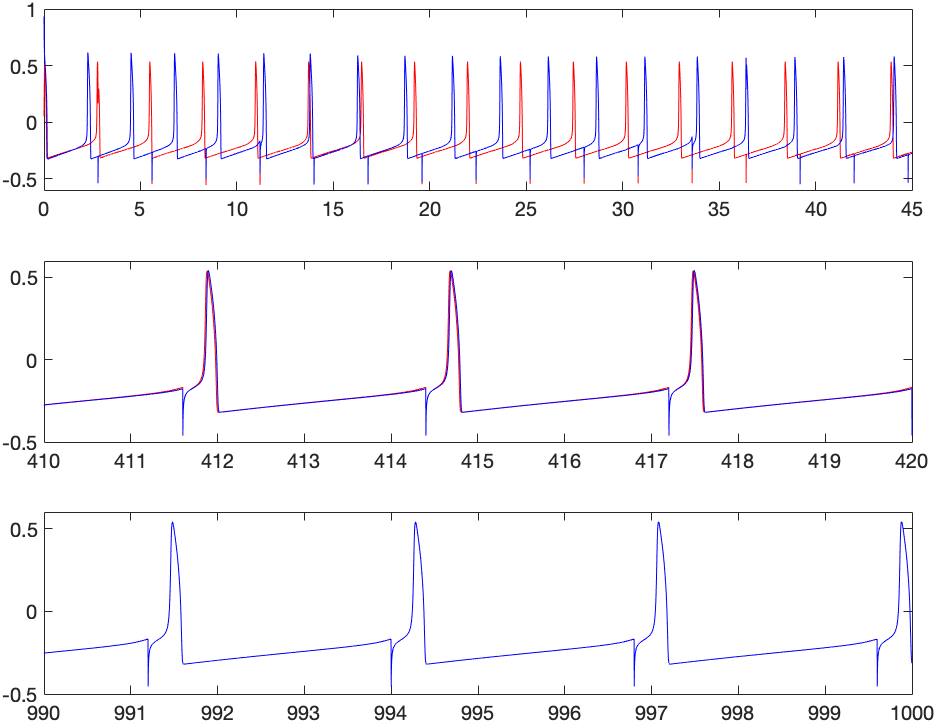}
\caption{Red: $y(\cdot)$ (which converges to $y_{\theta^{**}}^{**}(\cdot)$) and blue: $\hat{y}(\cdot)$.}
\label{fig:obs2}
\end{center}
\vspace{-.5cm}
\end{figure}

The linearization of~\eqref{eq:non_sys_obs} around the new reference output/state trajectory for the actual parameter $\theta^{**} \in \Theta$, ${\rm col}(y_{\theta^{**}}^{**}(\cdot), z_{\theta^{**}}^{**}(\cdot), \theta^{**})$, when the output injection $y(\cdot)$ in~\eqref{eq:non_sys_obs_c} is simply replaced by $y_{\theta^{**}}^{**}(\cdot)$, becomes
$$\begin{bmatrix} \dot{\delta \hat{y}} \\ \dot{\delta \hat{z}} \\ \dot{\delta \hat{\theta}}\end{bmatrix} = \begin{bmatrix} A_{\theta^{**}}^{**}(t) & \begin{bmatrix} h(y_{\theta^{**}}^{**}(t))^T \\ 0 \end{bmatrix} \\ \begin{bmatrix} -h(y_{\theta^{**}}^{**}(t)) & 0 \end{bmatrix} & 0 \end{bmatrix} \begin{bmatrix} \delta\hat{y} \\ \delta\hat{z} \\ \delta\hat{\theta}\end{bmatrix},$$
where $A_{\theta^{**}}^{**}(\cdot)$ is a time-varying matrix for the linearization of~\eqref{eq:non_sys_app} for the parameter $\theta^{**}$, which makes the differential Lyapunov function $\delta V_{\theta^{**}}$ decrease over a one period of time $T_{\theta^{**}}^{**}$.
Therefore, a differential Lyapunov function for the adaptive observer~\eqref{eq:non_sys_obs} given as ($\epsilon \ll 1$)
$$\delta W_\epsilon = \frac{1}{2}\delta\hat{y}^2 + \frac{1}{2}\delta\hat{z}^TP_{\theta^{**}}(t)\delta\hat{z} + \frac{1}{2}\delta\hat{\theta}^T\delta\hat{\theta} - \epsilon \delta\hat{\theta}h(y_{\theta^{**}}^{**}(t))\delta\hat{y}$$
also decreases, resulting in contraction.\footnote{Detailed analysis involving generalized eigenvalues and persistency of excitation conditions~\cite[Section 2.5]{sastry1990adaptive} on $h(y_{\theta^{**}}^{**}(\cdot))$ will be given in a follow-up paper.}
Notice that the solutions of the data-generating system~\eqref{eq:non_sys_app} with $\dot{\theta}=0$ are also solutions of the adaptive observer~\eqref{eq:non_sys_obs}, which guarantees parameter estimation convergence.
Figures~\ref{fig:obs1} and~\ref{fig:obs2} show the simulation result with system
\begin{align*}
0.02\dot{y} &= - 2 z(y + 0.7) + 0.15 + u - \begin{bmatrix} y + 0.4 \\ m_\infty(y)(y - 1) \end{bmatrix}^T\!\theta, \\
\tau(y)\dot{z} &= -z + z_\infty(y), \\
m_\infty(y) &= \text{sat}\left(0, 1, (-2y^3+0.9y^2+0.6y+0.068)/0.343\right), \\
\tau(y)  &= \text{sat}\left(0.2, 1, 0.2 + 40(0.25-y)\right),\\
z_\infty(y) &= \text{sat}\left(0, 1, (y + 0.17)/0.42\right),
\end{align*}
where $\text{sat}(a, b, s) = \max\{a, \min\{b, s\}\}$.
For the set $\Theta = [0.3, 0.7] \times [1.1, 1.9]$, an input trajectory $u^{**}(\cdot)$, given as the square wave of magnitude $-3$ having duration $0.002$ seconds and period $2.8$ seconds, induces contraction.\footnote{This control input perturbs the original trajectory to spend more time on the contractive region as noted in Remark~\ref{rem:cont}.}
The parameter used for simulation is $\theta^{**} = {\rm col}(0.5, 1.5)$, and the observer uses system copy for the variables $y$ and $z$, and the dynamics
\begin{align*}
\dot{\hat{\theta}} &= \begin{bmatrix} \hat{y}^2/2 + 0.4 \hat{y} - y^2/2 - 0.4 y\\ M_\infty(\hat{y}) - M_\infty(y) \end{bmatrix},
\end{align*}
for the parameters, where $M_\infty(y) = \int m_\infty(y)(y-1) dy$.

\section{Conclusion}\label{sec:conc}

In this note, we considered the problem of induced contraction by assuming that the inverse system is contracting.
Thus, an input perturbation design question has been simplified to an output perturbation design question, from which we have explored several approaches that even simplified the design.
Overall, the design of a new reference trajectory has been shown to have a connection with the time spent in the contractive region.
An application of this design has been found in system identification.
There are many avenues left open for research in this direction.
Analytical verification of the approaches is one of them, which will be provided in a follow-up paper.

Moreover, the present work paves the road for the stochastic version of this work: noise-induced synchronization, where a random input perturbation is designed so as to induce contraction.
In particular, the describing function method has a direct counterpart when a multi sine input perturbation is replaced by stochastic noise~\cite{vander1968multiple}.
The stochastic theory is appealing since it suggests that the environment can induce contraction. 
For instance, one could imagine that a neural model from neuroscience is non-contractive in vitro but becomes contractive in vivo, where noisy input perturbations are ubiquitous.

\bibliographystyle{IEEEtran}
\bibliography{Reference}

\begin{thebibliography}{10}
\providecommand{\url}[1]{#1}
\csname url@rmstyle\endcsname
\providecommand{\newblock}{\relax}
\providecommand{\bibinfo}[2]{#2}
\providecommand\BIBentrySTDinterwordspacing{\spaceskip=0pt\relax}
\providecommand\BIBentryALTinterwordstretchfactor{4}
\providecommand\BIBentryALTinterwordspacing{\spaceskip=\fontdimen2\font plus
\BIBentryALTinterwordstretchfactor\fontdimen3\font minus
  \fontdimen4\font\relax}
\providecommand\BIBforeignlanguage[2]{{%
\expandafter\ifx\csname l@#1\endcsname\relax
\typeout{** WARNING: IEEEtran.bst: No hyphenation pattern has been}%
\typeout{** loaded for the language `#1'. Using the pattern for}%
\typeout{** the default language instead.}%
\else
\language=\csname l@#1\endcsname
\fi
#2}}

\bibitem{kapitza1965dynamical}
P.~L. Kapitza, ``Dynamical stability of a pendulum when its point of suspension
  vibrates, and pendulum with a vibrating suspension,'' \emph{\emph{In}
  Collected papers of P.L. Kapitza. \emph{London, U.K.: Pergamon}}, vol.~2, pp.
  714--725, 1965.

\bibitem{ronsse2007rhythmic}
R.~Ronsse, P.~Lefevre, and R.~Sepulchre, ``Rhythmic feedback control of a blind
  planar juggler,'' \emph{IEEE Transactions on Robotics}, vol.~23, no.~4, pp.
  790--802, 2007.

\bibitem{russo2010global}
G.~Russo, M.~di~Bernardo, and E.~D. Sontag, ``Global entrainment of
  transcriptional systems to periodic inputs,'' \emph{PLoS Computational
  Biology}, vol.~6, no.~4, p. e1000739, 2010.

\bibitem{bryant1976spike}
H.~L. Bryant and J.~P. Segundo, ``Spike initiation by transmembrane current:
  {A} white-noise analysis,'' \emph{The Journal of Physiology}, vol. 260,
  no.~2, pp. 279--314, 1976.

\bibitem{mainen1995reliability}
Z.~F. Mainen and T.~J. Sejnowski, ``Reliability of spike timing in neocortical
  neurons,'' \emph{Science}, vol. 268, no. 5216, pp. 1503--1506, 1995.

\bibitem{zhou2003noise}
C.~Zhou and J.~Kurths, ``Noise-induced synchronization and coherence resonance
  of a {H}odgkin-{H}uxley model of thermally sensitive neurons,'' \emph{Chaos:
  An Interdisciplinary Journal of Nonlinear Science}, vol.~13, no.~1, pp.
  401--409, 2003.

\bibitem{bymes1992passivity}
C.~I. Byrnes, A.~Isidori, and J.~C. Willems, ``Passivity, feedback equivalence,
  and the global stabilization of minimum phase nonlinear systems,'' \emph{IEEE
  Transactions on Automatic Control}, vol.~36, no.~11, pp. 1288--1240, 1991.

\bibitem{lee2022induced}
J.~G. Lee and T.~B. Burghi, ``Funnel control by induced contraction,''
  \emph{accepted for presentation at the Proceedings of International Symposium
  on Mathematical Theory of Networks and Systems}, 2022.

\bibitem{boyd1985fading}
S.~Boyd and L.~O. Chua, ``Fading memory and the problem of approximating
  nonlinear operators with {V}olterra series,'' \emph{IEEE Transactions on
  Circuits and Systems}, vol.~32, no.~11, pp. 1150--1161, 1985.

\bibitem{lohmiller1998contraction}
W.~Lohmiller and J.-J.~E. Slotine, ``On contraction analysis for non-linear
  systems,'' \emph{Automatica}, vol.~34, no.~6, pp. 683--696, 1998.

\bibitem{strogatz2018nonlinear}
S.~H. Strogatz, \emph{Nonlinear dynamics and chaos: with applications to
  physics, biology, chemistry, and engineering}.\hskip 1em plus 0.5em minus
  0.4em\relax Perseus Books, 1994.

\bibitem{besanccon2000remarks}
G.~Besan{\c{c}}on, ``Remarks on nonlinear adaptive observer design,''
  \emph{Systems \& Control Letters}, vol.~41, no.~4, pp. 271--280, 2000.

\bibitem{farza2009adaptive}
M.~Farza, M.~M’Saad, T.~Maatoug, and M.~Kamoun, ``Adaptive observers for
  nonlinearly parameterized class of nonlinear systems,'' \emph{Automatica},
  vol.~45, no.~10, pp. 2292--2299, 2009.

\bibitem{burghi2021online}
T.~B. Burghi and R.~Sepulchre, ``Online estimation of biophysical neural
  networks,'' \emph{arXiv preprint arXiv:2111.02176}, 2021.

\bibitem{sastry1990adaptive}
S.~Sastry and M.~Bodson, \emph{Adaptive control: Stability, convergence, and
  robustness}.\hskip 1em plus 0.5em minus 0.4em\relax Prentice-Hall, 1989.

\bibitem{vander1968multiple}
A.~Gelb and W.~E. Vander~Velde, \emph{Multiple-input describing functions and
  nonlinear system design}.\hskip 1em plus 0.5em minus 0.4em\relax McGraw-Hill,
  1968.

\end{thebibliography}

\end{document}